\documentstyle[11pt,paspconf, epsf]{article}
\begin{document}
\title{The Correlation between Si III] $\lambda1892$/C III] $\lambda1909$ and 
Fe II $\lambda4500$/H$\beta$ in low redshift QSOs}
\author{Kentaro Aoki}
\affil{Astronomical Data Analysis Center, 
National Astronomical Observatory of Japan, 
2-21-1, Osawa, Mitaka, Tokyo 181-8588 Japan}
\author{Michitoshi Yoshida}
\affil{Okayama Astrophysical Observatory, National Astronomical Observatory of Japan,
Kamogata-cho, Asakuchi-gun, Okayama 719-0232, Japan}

\begin{abstract}
{\it HST} archival FOS spectra of 40 QSOs with z $\leq 0.5$ in the Bright Quasar Survey have been analyzed.
The spectra cover the region $\lambda\lambda1800-2000$ \AA~ in the QSOs' rest frames,
including the Al III $\lambda1859$, 
Si III] $\lambda1892$, C III] $\lambda1909$, and Fe III UV34
emission-lines.
We measured the flux of these UV emission-lines,
and analyzed the correlations among UV and optical (H$\beta$, Fe II, and [{O} III]) 
emission-line properties
as well as soft X-ray photon indices.
We found a significant correlation between Si III]/C III] 
and Fe II/H$\beta$.

Si III and C III have similar ionization potentials, 
but Si III] has one order magnitude larger critical density than C III].
Si III]/C III] is thus a density indicator and becomes larger when
density is higher.
The correlation between Si III]/C III] and Fe II/H$\beta$ 
indicates that optical Fe II becomes strong when the density of the broad line region becomes high.

Our correlation analysis shows that large Si III]/C III] associates with
weak [{O} III] $\lambda5007$, large soft X-ray photon index, and narrow
H$\beta$ width as well as with large Fe II/H$\beta$.  Our results support the
previous suggestions that the density of the broad line region gas and
the mass accretion rate govern this correlation. 

\end{abstract}

\keywords{galaxies: nuclei --- quasars: emission lines --- 
optical Fe II --- Si III]1892 --- C III]1909 --- ultraviolet: galaxies}

\section{INTRODUCTION}

Fe II emission-lines are one of the prominent features in the (rest) optical spectra of QSOs.
Numerous observational studies have investigated the optical Fe II emission, 
and several correlations between optical Fe II strength 
and other observational properties have been found 
(Boroson \& Green 1992 (BG92); Wang, Zhou, and Gao 1996;
Wang, Brinkmann, and Bergeron 1996 (WBB96); Laor et al. 1997a).

BG92  analyzed the spectral region including Fe II $\lambda4500$, H$\beta$, and 
[{O} III] $\lambda5007$ of all 87 QSOs with redshift less than 0.5 
in the Bright Quasar Survey (BQS) (Schmidt \& Green 1983).
They found a strong anticorrelation between the strength of Fe II and 
[O III] $\lambda5007$.
They also found that the asymmetry and width of the H$\beta$ line 
are associated with this anticorrelation.
The principal component analysis by BG92 revealed that the most significant eigenvector,
which accounts for 29\% of the variance, consists of this anticorrelation 
between optical Fe II and the [O III] emission-line.
BG92 called it Eigenvector 1.
They suggested that Eigenvector 1 was related to the fraction of the
Eddington luminosity.

Wang, Zhou, and Gao (1996) found that optical Fe II/H$\beta$ correlates with 
Si IV + O IV] $\lambda1400$/C IV $\lambda1549$ and 
C IV $\lambda1549$/Ly$\alpha$ in UV spectra of active galactic nuclei (AGN)
with z $< 0.211$ obtained with $IUE$.
They proposed three possible causes of this correlation: the inclination of the 
accretion disk, the spectrum of ionizing continuum, and the density of broad-line 
region (BLR) gas.

WBB96 also found that the soft X-ray photon index measured with ROSAT PSPC strongly correlates with 
the optical Fe II/H$\beta$ and H$\beta$ line widths in 86 AGN.
The AGNs with steep spectral slopes in the soft X-ray region have stronger optical Fe II and
narrower H$\beta$ broad lines.
WBB96 interpreted this to mean that the black hole mass and mass accretion rate 
govern the correlation between H$\beta$ width and the spectral slope
of soft X-rays.

Laor et al. (1997a) analyzed ROSAT PSPC spectra of the complete sample of BQS QSOs with z $\leq 0.400$ 
and $N_{H I}^{\rm Gal} < 1.9 \times 10^{20}$ cm$^{-2}$.
They found that the power law slope of the soft X-ray spectra correlates with optical 
Fe II/H$\beta$, H$\beta$ width and $L_{[O III]}$ in their sample.
The correlations they have found are similar to those found by WBB96.
The optical Fe II strength is related to the various properties of QSOs
such as the radiation from the accretion disk (soft X-ray spectral
slope),
the kinematical properties of BLR (H$\beta$ FWHM), 
and the luminosity of the narrow line region ([O III]
luminosity).
These properties arise from different structures ranging over more than four orders of
magnitude difference in scale in the nuclei,
from the inner radius of the accretion disk to a few hundred parsecs in the narrow line
region.
Study of these correlations will give us further understanding of 
QSOs.

Another property recently has been found to correlate with the optical Fe II
emission-line strength.
I Zw 1 (= PG 0050+124), which is known as a strong optical Fe II emitter (Phillips 1976), 
shows extremely strong Si III] $\lambda1892$ relative to C III] $\lambda1909$ (Laor et al. 1997b).
We therefore have begun to analyze the emission-feature around 1900~\AA~ in UV spectra of 
low-redshift (z $\leq 0.5$) QSOs in the BQS
obtained with the Faint Object Spectrograph (FOS) of the {\it Hubble Space Telescope} ({\it HST}).
The BQS is a UV-excess ($\ub < -0.44$) and magnitude limited ($B < 16.16$) survey.
The advantages of BQS QSOs are that they have been extensively observed over almost the entire 
electromagnetic waveband: radio, near-infrared, optical, and X-ray.
Especially, good S/N optical spectra of all 87 BQS QSOs with redshift less than 0.5 
have been reduced in the same manner and published by BG92.
The low redshift BQS sample includes only apparently bright QSOs, allowing good S/N UV spectra.
Wills et al. also have analyzed FOS spectra of the complete sample
of BQS QSOs.
Their results are presented in these proceedings.

Data and measurements of emission-lines are described in Section 2.
We present in Section 3 the correlations among our measurements of UV emission-lines and optical 
and soft X-ray properties from the literatures.
In section 4 we discuss the results and give interpretations.
We give concluding remarks in Section 5.
More details will be presented elsewhere (Aoki \& Yoshida 1998).

\section{DATA AND REDUCTION}

We have gathered {\it HST} archival FOS data of BQS QSOs with redshifts less than 0.5.
Two objects are omitted from the original list in Schmidt \& Green (1983),
as described in BG92, due to misclassification or a wrong redshift.
The spectra of 44 BQS QSOs with z $\leq 0.5$ cover $\lambda\lambda1800-2000$ \AA~ in the QSOs' rest frames.
They have been obtained with the G190H or G270H gratings (R $\sim$ 1300).
They were observed for various studies of intervening and intrinsic absorption-lines 
as well as emission-line properties of QSOs,
so this sample is not complete.
Five different  FOS apertures  were used to observe them, 
but the difference in resolution caused by different aperture size is less than 5\%.
If an object was observed more than twice within one or two days with similar exposure times, 
we combined data to improve S/N even if the aperture sizes are different. 
If observations of the same object were separated by several months or more, we chose the one with better S/N.
PG 1535+547 was observed with the spectropolarimetry mode,
so we used the total flux spectra of it.
Two objects, PG 0043+039 and PG 1700+518, are broad absorption line QSOs, for which it is difficult to do decompose the complex of emission lines around 1900~\AA.
They are excluded from our analysis.
The objects' names and redshifts are tabulated in Table 1.
\begin{table}[h]
\caption{Objects' list}
\begin{center}
\begin{tabular}{cccccccc}
\tableline
\tableline
PG & z & PG & z & PG & z & PG & z \\
\tableline
0003+158 & 0.450 & 1115+407 & 0.154 & 1351+640 & 0.087 & 1444+407 & 0.267 \\
0003+199 & 0.025 & 1116+215 & 0.177 & 1352+183 & 0.158 & 1512+370 & 0.371 \\
0026+129 & 0.142 & 1202+281 & 0.165 & 1402+261 & 0.164 & 1535+547 & 0.038 \\
0050+124 & 0.061 & 1211+143 & 0.085 & 1404+226 & 0.098 & 1545+210 & 0.266 \\
0052+251 & 0.155 & 1216+069 & 0.334 & 1411+442 & 0.089 & 1612+261 & 0.131 \\
0947+396 & 0.206 & 1226+023 & 0.158 & 1415+451 & 0.114 & 1626+554 & 0.133 \\
0953+414 & 0.239 & 1259+593 & 0.472 & 1416--129 & 0.129 & 1704+608 & 0.371 \\
1001+054 & 0.161 & 1302--102 & 0.286 & 1425+267 & 0.366 & 2112+059 & 0.466 \\
1049--005 & 0.357 & 1307+085 & 0.155 & 1427+480 & 0.221 & 2251+113 & 0.323 \\
1100+772 & 0.313 & 1309+355 & 0.184 & 1440+356 & 0.077 & 2308+098 & 0.432 \\
1114+445 & 0.144 & 1322+659 & 0.168 &&&&\\
\tableline
\end{tabular}
\end{center}
\end{table}
The spectra of six objects with strong Si III] $\lambda1892$ are shown in Fig. 1.
\begin{figure}
\plotone{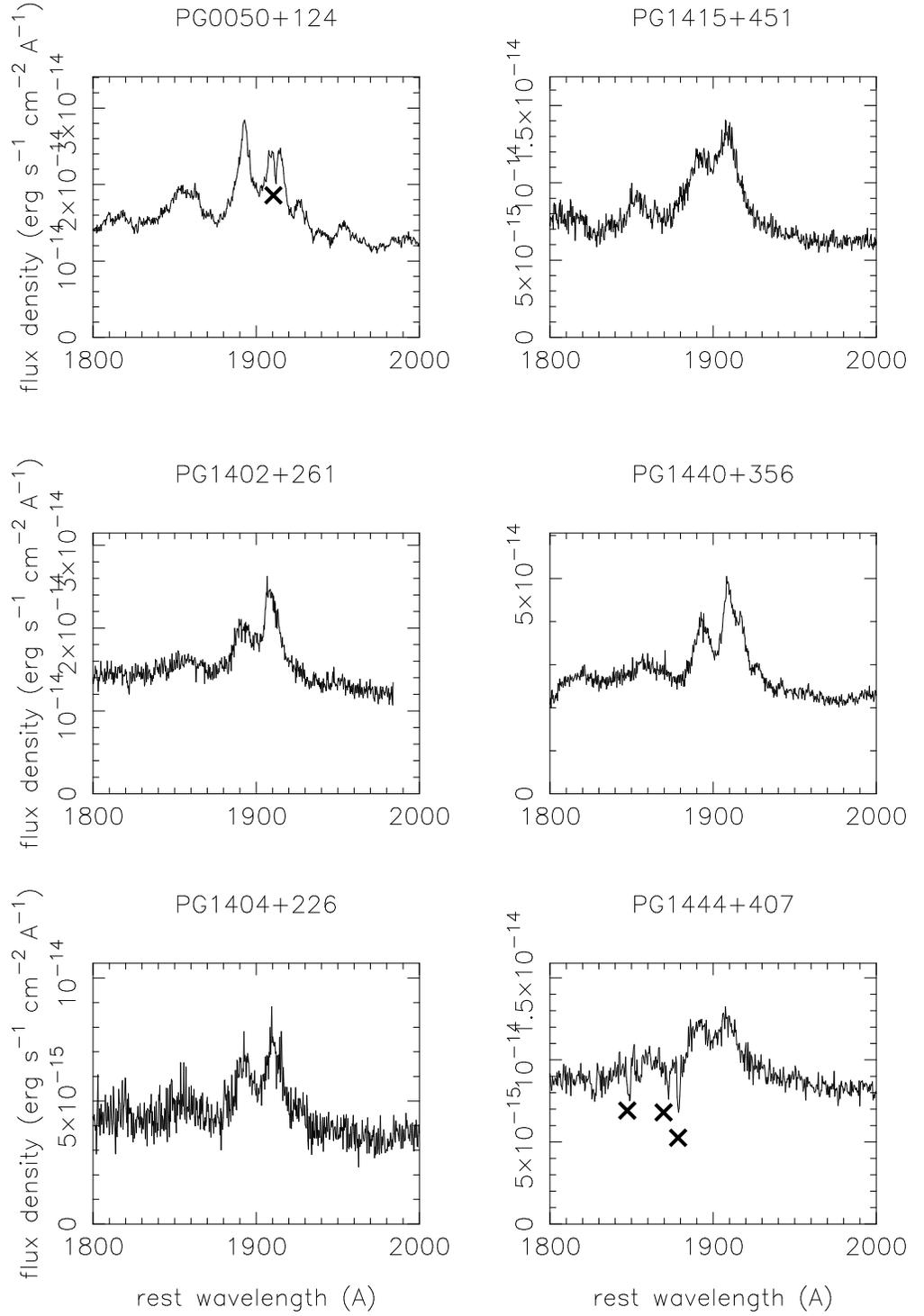}
\caption{The 1900 \AA~feature of six BQS QSOs with strong Si III] $\lambda1892$.
The ordinate is flux density (erg s$^{-1}$ cm$^{-2}$ \AA$^{-1}$),
the abscissa is rest wavelength (\AA).
Galactic absorption lines are marked.}
\end{figure}
They have been deredshifted using the measured peak wavelength of C III] $\lambda1909$.

The FOS data were pipeline processed at STScI, and flux and wavelength calibrated.
The line profiles were fitted using SPECFIT (Kriss 1994) in IRAF
\footnote{IRAF is distributed by the National Optical Astronomy 
Observatories, which are operated by the Association of Universities for
Research in Astronomy, Inc. under cooperative agreement with the National Science 
Foundation.}.
The 1900 \AA~ feature is supposed to be a blend of C III] $\lambda1909$, 
Si III] $\lambda1892$, and Fe III UV34 triplets $\lambda\lambda$1895, 1914, and 1926 
(see, for example, Baldwin et al. 1996; Laor et al. 1997b).
The 1900 \AA~feature may consist of five emission-lines, 
and Al III $\lambda\lambda$1855, 1863 are also near the 1900 \AA~ feature.
Limited S/N and overlapping Galactic intersteller absorption lines  
prevented automatic fitting of the spectra.
We therefore adopted a manual approach.
The acceptable fits were determined by eye rather than from $\chi^2$. 

PG 0026+129 was fitted by the Fe III triplets and a gaussian, 
and cannot be fitted by including corresponding components to 
Si III] $\lambda1892$ nor C III] $\lambda1909$.
PG 1259+593 is also excluded due to serious overlapping of a
Galactic interstellar medium absorption line. 
Thus these objects are excluded from following results and discussions.

In addition to our measurements, other observed properties have been
gathered from the literature.  Our interests especially focus on the Fe II
emission strength and the quantities which correlate with Fe II emission in
the previous studies (BG92; WBB96).  $R$ Fe II is the ratio of the
equivalent width of
the Fe II complex between $\lambda4434$ and $\lambda4684$ \AA~ 
to that of H$\beta$.
Peak $\lambda5007$ is the ratio of the peak height of the [{O} III] $\lambda5007$
to that of H$\beta$.
This quantity is difficult to understand, however, because it is strongly
anticorrelated with the Fe II complex strength (BG92).
H$\beta$ FWHM is the full width at half maximum intensity of the H$\beta$ emission line,
in km sec$^{-1}$.
$M_{\rm [O III]}$ is an [{O} III] $\lambda5007$ absolute magnitude 
defined as $M_{V} - 2.5$ log (EW [{O} III] $\lambda5007$).
$\Gamma_{\rm X}$ is the photon index of the soft X-ray spectra observed 
with the ROSAT PSPC. They have been gathered from WBB96.
WBB96 showed that this quantity correlates well with Fe II/H$\beta$.

\section{CORRELATIONS AMONG PROPERTIES}

Correlation coefficients ($r$) have been calculated among properties
taken from the literature and the UV line ratios which we measured: 
Si III] $\lambda1892$ / C III] $\lambda1909$ (Si III]/C III]),
Al III $\lambda1859$ / C III] $\lambda1909$ (Al III/C III]),
and Fe III UV34 / C III] $\lambda1909$ (Fe III/C III]).
We present the correlation matrix in Table 2.
\begin{table}[t]
\caption{Correlation Matrix}
\begin{center}
\begin{tabular}{ccccccccc}
\tableline
\tableline
 & $\frac{{\rm Al III}}{{\rm C III]}}$ & $\frac{{\rm Si III]}}{{\rm C III]}}$ & 
$\frac{{\rm Fe III}}{{\rm C III]}}$ & $R$ & Peak &
H$\beta$ & $M_{\rm{[O III]}}$ & $\Gamma_{\rm X}$ \\
 & & & & Fe II & $\lambda5007$ & FWHM & & \\
\tableline
Al III/C III] & 1 & & & & & & & \\
Si III]/C III] & 0.69 & 1 & & & & & & \\
Fe III/C III] & 0.67 & 0.57 & 1 & & & & & \\
$R$ Fe II  &0.66& 0.72 & 0.68 & 1 & & & & \\
Peak 5007 &-0.34& -0.49& -0.39& -0.65 & 1 & & & \\
H$\beta$ FWHM & -0.37& -0.47& -0.55 & -0.68& 0.70& 1 & & \\
M$_{\rm [O III]}$ & 0.37& 0.55& 0.47&  0.71& -0.77& -0.69& 1 & \\
$\Gamma_{\rm x}$ & 0.35& 0.49& 0.61& 0.60& -0.48& -0.70& 0.59& 1 \\
\tableline
\end{tabular}
\end{center}
\end{table}
Each correlation coefficient is calculated using only the objects
for which both values are given; 
all 40 objects are used except that correlations involving the photon index ($\Gamma_{\rm X}$) used only 
the 34 objects for which that quantity was measured.

There are many strong correlations in Table 2.
Five absolute values of correlation coefficients are larger than 0.7, 
and nine absolute values are between 0.6 and 0.7.
The measurement of Fe II strength ($R$ Fe II) 
anticorrelates with [O III] strength (Peak $\lambda5007$ and $M_{\rm [O III]}$)
and H$\beta$ FWHM in our sample, as BG92 had already found.
The correlation coefficients among them in our sample are larger than those in BG92
because the size of our sample (40 objects) is smaller than half that of BG92 (87 objects).

The significant correlations between $\Gamma_{\rm X}$ and 
$R$ Fe II and between $\Gamma_{\rm X}$ and H$\beta$ FWHM 
which exist in our sample previously have been  reported in WBB96.
$\Gamma_{\rm X}$ correlates with $M_{\rm [O III]}$.
Laor et al. (1997a) previously found a similar correlation between 
the power-law slope of soft X-ray spectra and [O III] luminosity
in their BQS sample. 

The correlation coefficient between $R$ Fe II and the UV line ratios we measured, 
Si III]/C III], Al III/C III], and Fe III/C III], are large (Table 2).
The correlation coefficient between Si III]/C III] and $R$ Fe II
is the largest one except for that between Peak $\lambda5007$ and 
$M_{\rm [O III]}$, which is between essentially the same properties.
It is clearly seen that there are two ``zone of avoidance'' 
which are right lower half and left upper half in Fig. 2(a).
\begin{figure}
\plotone{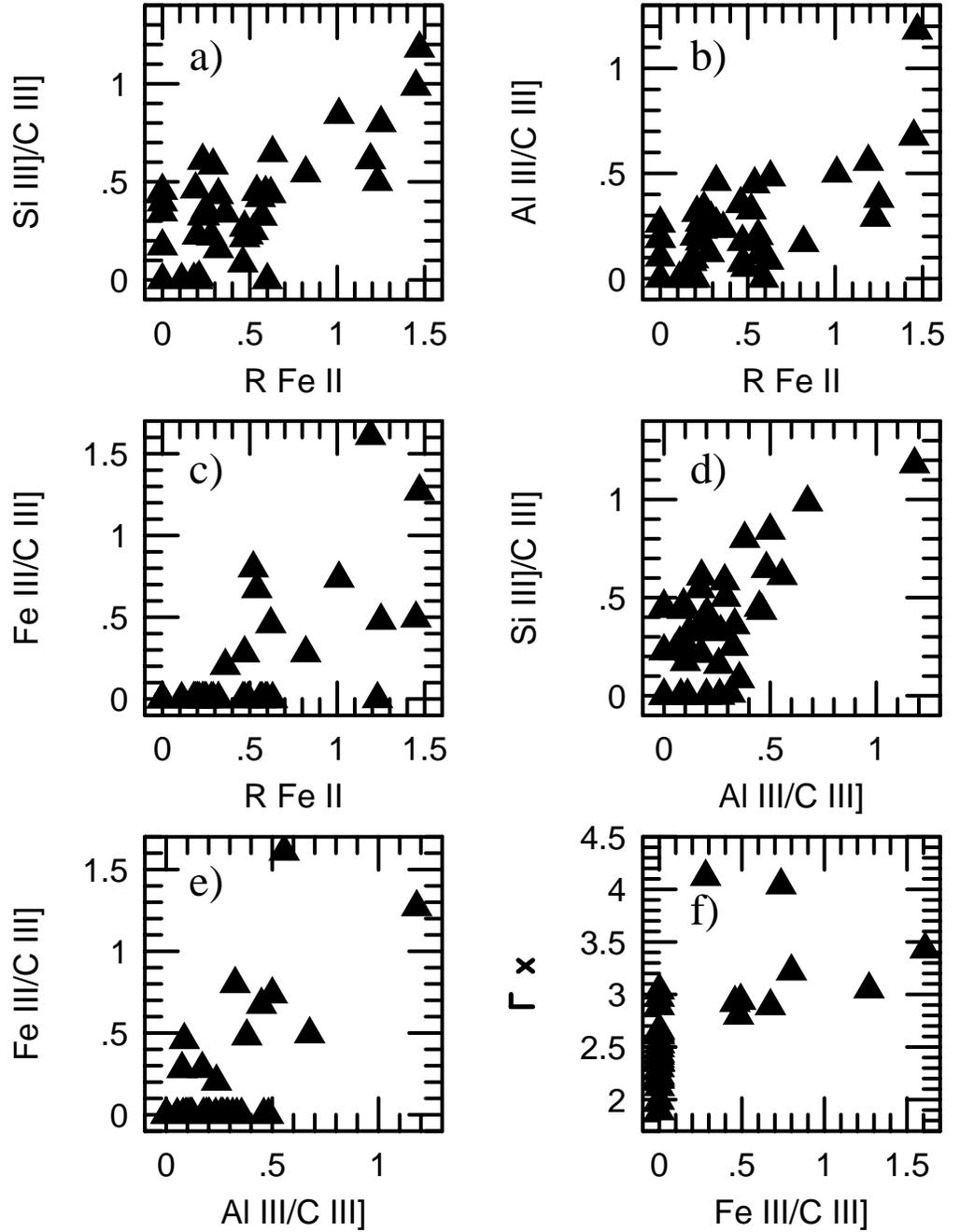}
\caption{a) The intensity ratio of Si III] $\lambda1892$ to C III] $\lambda1909$
(Si III]/C III]) is
plotted against the ratio of equivalent width of Fe II $\lambda4500$ 
to that of H$\beta$ ($R$ Fe II).
b) The intensity ratio of Al III $\lambda1859$ to C III] $\lambda1909$
(Al III/C III]) is
plotted against $R$ Fe II.
c) The intensity ratio of Fe III UV34 to C III] $\lambda1909$
(Fe III/C III]) is
plotted against $R$ Fe II.
d) Si III]/C III] is
plotted against Al III/C III].
e) Fe III/C III] is
plotted against Al III/C III].
f) The photon index of soft X-ray spectra ($\Gamma_{\rm X}$)
is plotted against Fe III/C III].}
\end{figure}
Si III]/C III] does correlate with $R$ Fe II, although
recognizing the uncertainty due to the small size of our sample.
The correlations between Al III/C III] and $R$ Fe II (Fig. 2(b))
and between Fe III/C III] and $R$ Fe II (Fig. 2(c)) 
also have similar uncertainty
due to the small size of the sample with $R$ Fe II $\geq 0.6$.

Among UV line ratios, Al III/C III] correlates with 
Si III]/C III] (Fig. 2(d)) and Fe III/C III] (Fig. 2(e)).
The scarcity of objects with Al III/C III] $\geq 0.4$
weakens the confidence in these correlations.

The correlation coefficient between Fe III/C III] and $\Gamma_{\rm X}$
is large ($r=0.61$) (Table 2).
The distribution of data, however, indicates there are two groups 
(Fig.  2(f)).
The average $\Gamma_{\rm X}$ of objects with Fe III/C III] $> 0$ 
and ones with Fe III/C III] $=0$ are $3.3\pm0.47$ and $2.5\pm0.28$,
respectively.
Strong Fe III emitters have steep soft X-ray spectra.

\section{DISCUSSION}

\subsection{The Intensity ratio Si III]/C III] as a density indicator}

Si III] 1892 has one order of magnitude larger critical density 
(n$_{cr} = 1.1 \times 10^{11}$ cm$^{-3}$) than C III] 1909
(n$_{cr} =  5 \times 10^{9}$ cm$^{-3}$).
The ionization potentials for making and destroying Si III and C III are
16.3 and 33.5, and 24.4 and 47.9 eV, respectively.
This line intensity ratio, therefore, is sensitive to the density of the gas.
Using the results of BLR photoionization calculations made by 
Korista et al. (1997), we investigated the dependence of Si III]/C III]
on the hydrogen density, n(H), and the ionization parameter, $U$(H).
This photoionization calculation assumes that 
the hydrogen column density is $10^{23}$ cm$^{-2}$,
the continuum spectral energy distribution peaks around 44 eV,
$\alpha_{\rm OX}$ is -1.40, and the Solar abundance (Korista et al. 1997).
Si III]/C III] is mainly a function of n(H) 
up to n(H) $\sim 10^{11}$ cm$^{-3}$ over a wide range of $U$(H) 
($-1.5 >$ log $U$(H) $> -3.5$).
When n(H) becames denser than $10^{11}$ cm$^{-3}$, Si III]/C III] remains
constant.
These behaviors are understood to be the result of the critical densities of Si III] and C III]. 

Si III]/C III] correlates with Al III/C III]
although there is an uncertainty due to the small sample size as described in Section 3.
Al III 1859 is a permitted line, and 
the ionization potentials for making and destroying Al III are
18.8 and 28.4 eV, respectively.
According to the same photoionization calculations by 
Korista et al. (1997), 
Al III/C III] increases with increasing n(H) when 
log $U$(H) $< -1$.
Al III/C III] and Si III]/C III] 
increase with the hydrogen density up to n(H) $\sim 10^{11}$ cm$^{-3}$.
The density variation of the BLR is the prime mover of the correlations between 
Al III/C III] and Si III]/C III].
Laor et al. (1997b) also interpreted a large Si III]/C III] of I Zw 1 
as being due to denser BLR gas.

Fe III/C III] also correlates with Al III/C III] ($r = 0.67$), 
and correlates with Si III]/C III] ($r = 0.57$).
The ionization potential for making and destroying Fe III is 16.2 and 30.7 eV, 
which are similar to those of Si III and Al III.
Fe III UV34 is a permitted line, thus its strength correlates more significantly 
with that of Al III 1859 than Si III] 1892 because of less collisional deexcitation.
Larger Fe III/C III] also indicates higher density.

\subsection{Interpretations of Eigenvector 1}

Si III]/C III] correlates significantly with $R$ Fe II ($r = 0.72$).
It also correlates with $M_{\rm [O III]}$ ($r = 0.55$),
Peak $\lambda5007$ ($r = -0.49$), and H$\beta$ FWHM ($r = -0.47$). 
Si III]/C III] therefore associates with Eigenvector 1 found by BG92.
Si III]/C III] also correlates with $\Gamma_{\rm X}$ ($r = 0.49$).
$\Gamma_{\rm X}$ has previously been found to correlate strongly with H$\beta$ width (WBB96; Laor et al. 1997a),
with $R$ Fe II (WBB96; Laor et al. 1997a), with [{O} III] luminosity (Laor et al. 1997a)
and with Peak $\lambda5007$ (Laor et al. 1997a).
These same correlations with $\Gamma_{\rm X}$ appear in our sample (Table 2).
$\Gamma_{\rm X}$ associates with Eigenvector 1 as well as with Si III]/C III].

The correlation of Si III]/C III] with $R$ Fe II is interpreted as being due to the existence of dense gas in the BLR.
The ionizing potentials of Fe II and Fe III are 7.9 eV and 16.2 eV, respectively.
Fe II is emitted from the partially ionized region of hydrogen.
High density gas can guarantee large optical depth which is necessary for development of a
partially ionized region.
Models calculated with high density have reproduced strong optical 
Fe II emission (Wills, Netzer, \& Wills 1985).

An additional hint that $R$ Fe II depends on the density of the BLR gas comes from
the significant correlation of Si IV + O IV] $\lambda1400$/C IV $\lambda1549$
with $R$ Fe II (Wang, Zhou \& Gao 1996).
Wang, Zhou \& Gao (1996) proposed high density gas as one of the causes. 
Si IV $\lambda1397$ becomes even stronger at densities higher than $10^{11}$ cm$^{-3}$
where C IV $\lambda1549$ is thermalized (Korista et al. 1997).

Laor et al. (1997a) interpreted the correlation of soft X-ray slope with H$\beta$
emission-line width as being due to a change in $L/L_{\rm Edd}$, the ratio of luminosity 
to Eddington luminosity of given mass.
One ground for that interpretation is that Galactic black hole candidates (GBHC) show steeper x-ray slopes 
when they become brighter, that is, when the mass accretion rate becomes larger (eg., Ebisawa et al. 1994)
although the physical mechanism controlling both the luminosity and the spectrum has not been understood.
The other ground is that $L/L_{\rm Edd} \propto v^{-2} L^{1/2} $ ($v$: velocity of BLR gas) 
assuming that the BLR is virialized and that
the size of the BLR is determined by the luminosity, $R \propto L^{1/2}$.
This scaling is based on the results of reverberation mapping (Kaspi et al. 1996)
and on theoretical predictions of the radius of dust sublimation (Laor \& Draine 1993; Netzer \& Laor 1994).
Larger $L/L_{\rm Edd}$ at a given black hole mass means a larger mass accretion rate which may reflect the availability of
more fuel.
More fuel may relate to a gas rich environment and dense BLR gas.

An alternative interpretation is that this is an inclination effect.
Puchnarewicz et al. (1992) suggested that a face-on view of an accretion disk will produce a
narrow H$\beta$ width and a strong soft x-ray excess if the BLR gas clouds are localized in a plane 
and coplanar with an accretion disk.
They favored a geometrically thick disk which would produce a strong soft x-ray excess when viewed face-on.
In PG 0050+124, however, a steep x-ray spectrum is produced without a strong 
soft x-ray excess. 
It is necessary to obtain x-ray spectra of more BQS QSOs.

Why is [{O} III] $\lambda5007$ weak for the QSOs with narrow
H$\beta$, steep soft x-ray spectra, and dense BLR gas?  Two possible
reasons can be considered.  One is the possible lack of an appropriate
gas ($10^{2}-10^{6}$ cm$^{-3}$ and $10^{4}$ K) for producing the [{O}
III] $\lambda5007$ emission-line.  A gas-rich environment which leads
to a dense BLR may suppress forbidden emission-lines such as [{O} III]
due to relatively high density.  Or gas may be removed as the results
of infall into the BLR or outflow driven by supernovae explosions.  The
other reason is that the excitation might be too low for very much [{O}
III] to be emitted.  In this case relatively strong [{O} II] emission
will be expected, but [{O} II] $\lambda3727$ was not detected in PG
0050+124 (Laor et al. 1997b).

\section{CONCLUSION}

We have analyzed 40 FOS spectra of BQS QSOs with redshift less than
0.5.  The fluxes of the Al III, Si III], C III] and Fe III
emission-lines around 1900 \AA~ were measured.  We calculated
correlation coefficients among the intensity ratios of these lines, the
soft x-ray photon index and several optical emission-line properties
(such as the ratio of Fe II to H$\beta$) taken from the literature.  A
significant correlation of Si III]/C III] with $R$ Fe II is found.  Si
III]/C III] also is associated with Eigenvector 1 found by BG92.  Si
III]/C III] is a density indicator and positively correlates with
density up to $10^{11}$ cm$^{-3}$.  These results support the
suggestion that the density variation relates
with $L/L_{\rm Edd}$, that is, with the mass accretion rate, and that this is one
of the causes of Eigenvector 1 found by BG92.

Many observational grounds for our interpretation depend on only one
BQS QSO, PG 0050+124.  Observations of other QSOs are necessary to
study relationships associated with Eigenvector 1 and the cause for
it.  X-ray spectroscopy is needed to study the radiation mechanism
producing the x-rays.  In the future we can hope to discuss the x-ray
continuum not only in terms of slope but also of radiation mechanisms.

\acknowledgments
We thank the Editor, Jack Baldwin, for his careful reading, 
helpful comments and many helps with the English.
This research has been carried out using the facilities at the
Astronomical Data Analysis Center, National Astronomical Observatory of
Japan, which is an inter-university research institute operated by the
Ministry of Education, Science, Culture and Sports.

\end{document}